\documentclass[aps,twocolumn,showpacs,showkeywords]{revtex4}
\newcommand{\ket}[1]{\left|#1\right\rangle}
\usepackage{graphicx}
\usepackage{setspace}

\begin{document}

\title{The excited state structure of the nitrogen-vacancy center in diamond}
\author{Ph Tamarat$^{1}$}
\author{N. B. Manson$^{2}$}
\author{R. L. McMurtie$^{2}$}
\author{A Nitsovtsev$^{3}$}
\author{C. Santori$^{4}$}
\author{P. Neumann$^{1}$}
\author{T. Gaebel$^{1}$}
\author{F. Jelezko$^{1}$}
\author{P. Hemmer$^{5}$}
\author{J. Wrachtrup$^{1}$}
\affiliation{1. Physikalisches Institut, Universit\"{a}t Stuttgart, 70550 Stuttgart, Germany}
\affiliation{2. Laser Physics Center, RSPhysSE, Australian National University, Canberra, ACT 0200,
Australia} \affiliation{3. Stepanov Institute of Physics, National Academy of Sciences of Belarus,
Minsk,220072, Belarus} \affiliation{4. Hewlett-Packard Laboratories, 1501 Page Mill Rd., Palo Alto,
CA 94304,USA} \affiliation{5. Texas A $\&$ M University, College Station, TX, 77843,USA}
\date{\today}

\begin{abstract}
Optical and microwave double resonance techniques are used to obtain the excited state structure of
single nitrogen-vacancy centers in diamond. The excited state is an orbital doublet and it is shown
that it can be split and associated transition strengths varied by external electric fields and by
strain. A group theoretical model is developed. It gives a good account of the observations and
contributes to an improved understanding of the electronic structure of the center. The findings are
important for quantum information processing and other applications of the center.
\end{abstract}


\maketitle

Impurity spins in solids are appealing for quantum information processing and information storage and
the nitrogen-vacancy center (NV) in diamond is a defect center with a spin ground state that offers
great promise \cite{EITQ,C13Q,2qub,Quan}. It has been shown that it is possible to optically detect
the center at single site level \cite{single site}. Broadband optical excitation causes
preferentially population of one of the ground state levels \cite{spin polarization} providing a
simple way of initializing the quantum system. More significantly it has been shown that using one
laser excitation it is possible to read-out the spin state of a single site \cite{Spin,Sing} and
using two-laser excitation to manipulated spins of a single site \cite{EIT}. Conceivably these
capabilities can offer the possibility of all optical control of the electron spin. What is not clear
is whether the processes can be achieved simultaneously with the same NV center and this is vital if
optical control is to be realized. The readout requires an almost perfectly cyclic spin-conserving
transition \cite{Spin,Quan} and the manipulation requires a non-spin conserving or $\Lambda$-type
transition \cite{smallEIT}. These are contrasting situations and the observations may have involved
centers in very different environments. It is not known how a cyclic or $\Lambda$-type transitions is
obtained and this is a consequence of having a poor understanding of the excited state of the NV
center. The present work has been directed at improving this situation. Bichromatic laser and
microwave optical double resonance measurements are made to obtain the excited state structure of
single NV centers and variations with external electric fields are also obtained. The observations
are compared with a group theoretical model and good correspondence is obtained. The model provides
the necessary insight to the excited state structure and enables the immediate questions regarding
the transition types to be answered and as to how the transitions maybe optimized. However, the work
has wider implication as it leads to an excellent understanding of the electronic structure of the NV
center and will enable more satisfactory development of all NV applications.

The NV defect in diamond comprises of a nitrogen atom at a lattice site next to a carbon vacancy
giving a center with C$_{3v}$ symmetry. The center gives an allowed transition between an orbital
A$_{2}$ ground state and an orbital E excited state \cite{Davi}. Both the ground and excited states
are spin triplets (S = 1) \cite{Redd}. The spin levels in the $^{3}$A$_{2}$ ground state are split by
2.88 GHz into a spin singlet S$_{z}$ and a spin doublet (S$_{x}$,S$_{y}$). From hole burning
measurements \cite{Redd,smallEIT} it is known that the $^{3}$E excited state is likewise split by
several GHz but the details have not been determined. The $^{3}$A$_{2}$ - $^{3}$E line widths are of
the order of several MHz \cite{Spin,Star} and it should be possible to establish the excited state
structure from an excitation spectrum. However, excitation spectrum of single NV centers have given
only one line \cite{Sing, Spin, Star} rather than the several anticipated. The reason for the smaller
number is attributed to differences in the axis of quantization in ground and excited state such that
with one laser for some of the transitions the center is immediately driven out of resonance by
non-spin-conserving electronic transitions. Therefore, the complete $^{3}$A$_{2}$ - $^{3}$E fine
structure cannot be obtained with one-laser but can be obtained with two and this is the approach
used here.

A laser is modulated to give a beam with a central frequency and sidebands at $\pm$ 2.88 GHz (Fig
1(a)). The laser is tuned to the $^{3}$A$_{2}$ - $^{3}$E zero-phonon line at 637 nm and focused to
within a type IIa diamond sample at 4.2K. The vibronic emission at 650 - 850 nm is detected in a
confocal arrangement to obtain the dual-laser excitation spectrum of single N-V centers \cite{Star}.
Using modulation, spectral responses are obtained at several wavelengths whereas there are no
equivalent responses in the absence of the modulation. An illustration of this situation is given by
the central and lower spectra in Fig 1 (b). The laser-sideband separation matches the ground spin
splitting and responses are obtained when two of the optical frequencies become resonant with
transitions from the ground state spin levels to a common excited state level in a $\Lambda$-type
scheme. As the ground state levels are associated with distinct spins states, S$_{z}$ and
(S$_{x}$,S$_{y}$), there will only be allowed transitions when the excited state has a mixed spin
character. The mixing occurs most efficiently between adjacent excited states and both of these mixed
states in general will satisfy the conditions for emission. Hence, there will be two features in the
sideband (SB) excitation spectrum. This is consistent with observation and in Fig 1(b) and Fig 2(b)
where the two features are separated by 0.32 GHz. The features are assigned to the 1-2 and 3-4
transitions in the energy level scheme in Fig 2(a). The spectral features are repeated in the SB
spectrum at an interval of 2.88 GHz when the laser and the reverse sideband are resonant with the
transitions.

\begin{figure}
\includegraphics[width=0.45\textwidth]{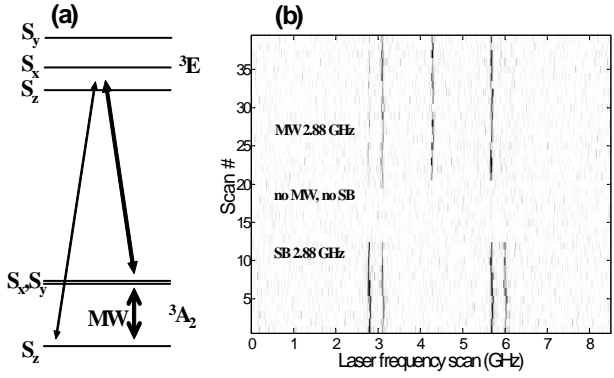}
\caption{(a) Energy levels and laser excitation fields. (b)Spectral trails of the same NV center when
irradiated with a laser plus sidebands at $\pm$2.88 GHz (lowest), with single laser (middle) and with
a laser plus microwaves at 2.88 GHz (highest).} \label{Fig1}
\end{figure}

\begin{figure}
\includegraphics[width=0.45\textwidth]{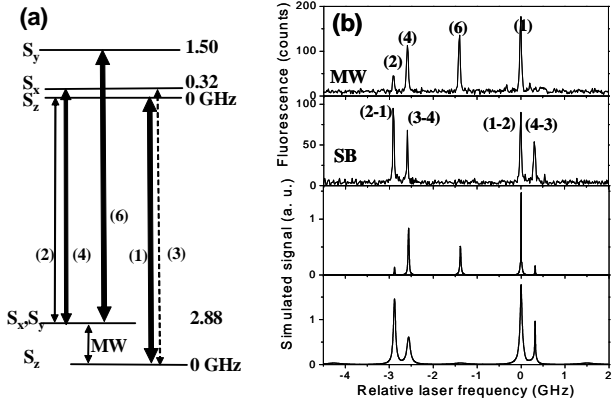}
\caption{(a) Energy levels and assignment of transitions. (b) Spectra taken using single laser plus
microwaves at $\pm$2.88 GHz (top) and using lasers plus sidebands (second top). The lower traces are
obtained from simulations including optical and microwave fields for a seven level model. The
slightly larger line widths obtained in the experiment are attributed to laser frequency drifts
during the experiment.} \label{Fig2}
\end{figure}

In a second approach microwaves (MW) are applied resonant with the ground state transition at 2.88
GHz to maintain a time-averaged population within each of the ground state spin levels. A single
laser is then swept in frequency to give a more conventional excitation spectrum. Such a MW spectrum
is shown in the upper traces in Fig 1(b) and Fig 2(b).  The transitions, 1, 2 and 4 to the two
excited state levels separated by 0.32 GHz are observed.  An additional transition, 6, is observed
and interpreted to be to an excited state level at 1.5 GHz  (Fig 2(a)).

The measurements in Fig 1 and Fig 2 are interpreted in terms of transitions between three spin levels
S$_{z}$, S$_{x}$, S$_{y}$ in both the ground and the excited state each considered to be orbital
singlets. The interpretation is consistent with that given in previous publications \cite{NVre}. A
simulation of the spectra using the six spin levels (plus an intermediate singlet level) is given in
the lower traces in Fig 2(b) and gives reasonable agreement with the experiments. There is still the
need to account for the behavior where one excitation line is obtained with a single laser. This line
is attributed to a different orbital level and, hence, the two spectra assigned to the two orbital
components of a strain-split $^{3}$E excited state. If this interpretation is correct all the
features should be observable for a single center and this is the situation obtained in Fig 3(a). The
higher energy features can be obtained with a single laser and the lower features only when using a
modulated laser beam. The features in the upper and lower spectra exhibit correlated jumps and have
orthogonal polarization supporting their assignment to the same defect. The modulation used to obtain
both spectral types in a single scan results in the repetition after 2.88 GHz.

\begin{figure}
\includegraphics[width=0.45\textwidth]{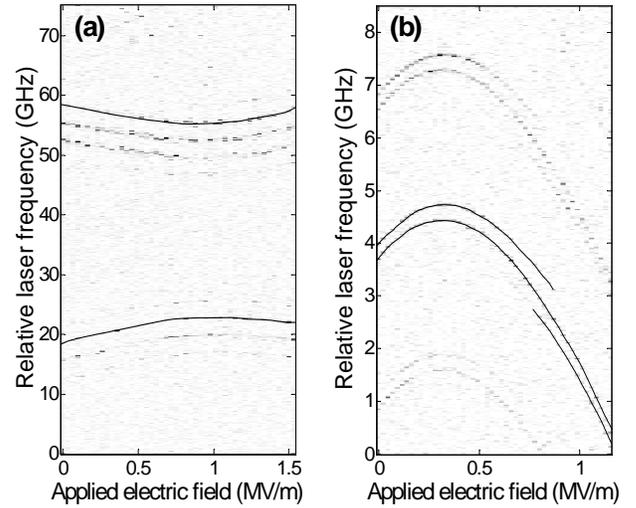}
\caption{$^{3}$A$_{2}$ - $^{3}$E excitation spectra as a function of external electric field. (a) Two
sets of excitation lines associated with a single NV center obtained using laser beam modulated at
2.88 GHz. (b) The variation of the excitation spectra of a different NV center as a function of
electric field. In both figures some of the traces have been manually enhanced.} \label{Fig3}
\end{figure}

Electrodes were deposited on the $<$001$>$ surface of the diamond to enable electric fields of up to
3 MV m$^{-1}$ to be applied. This field has components parallel, F$_{z}$, and perpendicular,
(F$_{x}$,F$_{y}$), to the axis of the NV center (x is defined to be in the C$_{3V}$ reflection
plane). For the latter spectrum (Fig 3(a)) the electric field causes the higher energy features to be
displaced with a curvature to higher energy. This is a characteristic that has been reported
previously \cite{Star}. The lower energy features have the reverse behavior with a curvature to lower
energy. The electric field also causes changes to the fine structure and typical variations are shown
in Fig 3(b). The changes in separations and intensities are consistent with two avoided crossings of
the spin levels occurring as a function of electric field

The observations can be explained by considering the effects on the $^{3}$E state with the
Hamiltonian:

H   =   H$_{0}$  +  H$_{so}$  + H$_{elec}$  +  H$_{str}$

H$_{0}$ gives the energy of the state. H$_{so}$ is the spin orbit interaction and H$_{elec}$ and
H$_{str}$ are perturbations due to electric fields and strain, respectively.  In C$_{3v}$ symmetry
the S$_{z}$ spin wave-function transforming as A$_{2}$ and the associated $^{3}$E spin-orbit state
has A$_{2}\otimes$ E = E symmetry. Spins S$_{x}$, S$_{y}$ transform as rows of an E representation
and the spin-orbit states have E $\otimes$ E = A$_{1}$, A$_{2}$, E symmetries. The four spin-orbit
states are affected by spin-orbit:

H$_{so}$ = $\lambda$(L$_{z}$S$_{z}$ + L$_{x}$S$_{x}$ + L$_{y}$S$_{y}$)

First consider the effect of the transverse spin-orbit interaction. The spin operators S$_{x}$,
S$_{y}$ can in principle lead to mixing of spin states but the associated orbital coefficient
$\lambda$$<$E$||$L$_{xy} ||$E$>$ = \textit{e} is small. Initially this value is taken to be zero such
that there is no S$_{z}$ - (S$_{x}$,S$_{y}$) spin mixing and the wavefunctions are determined by
symmetry \cite{NVre}. Axial spin-orbit interaction $\lambda$L$_{z}$S$_{z}$ will not affect the
($^{3}$E)E state associated with S$_{z}$ spin but will displace the states associated with
S$_{x}$,S$_{y}$ spins (($^{3}$E)A$_{1}$,A$_{2}$) and ($^{3}$E)E by $\lambda$$<$E$||$L$_{xy}||$E$>$ =
\textit{a} and - \textit{a}, respectively. From a six electron (two hole) model \textit{a} is
considered to be positive \cite{NVre}.

The electronic model of the NV center anticipates intermediate $^{1}$A$_{1}$ and $^{1}$E singlet
levels \cite{NVre}. Spin-orbit interaction with the $^{1}$A$_{1}$ will cause the ($^{3}$E)A$_{1}$
state to be displaced to higher energy by \textit{b} and similarly that with the $^{1}$E singlet
level can cause one of the ($^{3}$E)E states (lower) to be shifted up in energy by \textit{c}. This
latter shift could also arise from spin-spin interaction. \textit{a, b} and \textit{c} are treated as
semi-empirical parameters and are estimated from experiments. Values used here are \textit{a} = 4.4
GHz, \textit{b} = 2 GHz, \textit{c} = 1 GHz and these values determine the zero field energy levels
in Fig 4(a).

The electric field perturbation is given by

H$_{elec}$ = F$_{z}$D$_{z}$ + F$_{x}$D$_{x}$ + F$_{y}$D$_{y}$

where F$_{i}$ is the electric field in direction i = z, x, y. D$_{i}$ is the associated orbital
operator. The first term F$_{z}$D$_{z}$ will simply give rise to linear shift of all levels and not
considered further. The transverse terms will cause linear splittings of the orbitally degenerate E
state. The splitting is independent of field direction in xy plane and is given by $\pm$ \textit{d}F
= $\pm$ \textit{d} $\sqrt{(F_{x}^{2} + F_{y}^{2})}$ where \textit{d}= $<$E$||$D$||$E$>$. From
previous work \textit{d} is of order of 6 GHz /MVm$^{-1}$ \cite{Star}. The orbital eigenstates vary
with direction of the electric field and for a F$_{x}$ field are $\ket{E_{x}}$ and $\ket{E_{y}}$, and
a F$_{y}$ field $\ket{E_{x}} \pm \ket{E_{y}}$. The variation in the energy of the $^{3}$E spin-orbit
states for a field F$_{y}$ applied along the y direction is calculated and shown in Fig 4(a). The
dominant effect is the linear splitting of the orbital components.

\begin{figure}
\includegraphics[width=0.45\textwidth]{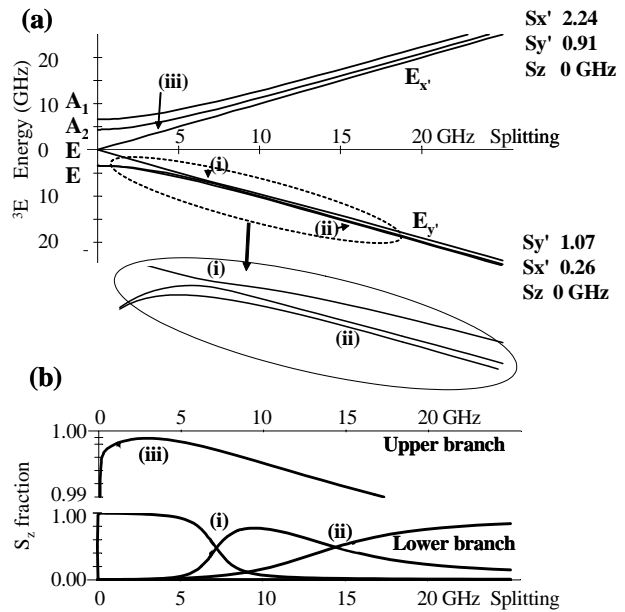}
\caption{(a) Energy levels of the excited $^{3}$E excited state as a function of the orbital
splitting calculated for a F$_{y}$ field. At zero field the levels are determined by the spin-orbit
parameters \textit{a}, \textit{b}, and \textit{c} given in text. The wavefunctions $\ket{E_{x'(y')}}$
= $\ket{E_{x}} \pm \ket{E_{y}}$ and $\ket{S_{x'(y')}}$ = $\ket{S_{x}} \pm \ket{S_{y}}$. (b)
Eigenstates giving fraction of S$_{z}$ within each state. In the upper case there is only the one
state with significant S$_{z}$ component. Special features are labelled (i) - (iii).} \label{Fig4}
\end{figure}

In the experiment a quadratic dependence is obtained and is due to the presence of strain. This
differs from the interpretation given in reference [13]. The terms within H$_{str}$ transform as
A$_{1}$ and E irreducible representations, the same as those for the electric field. Therefore, the
inherent strain has the same effect as a weak electric field and the components along z, x and y
directions are denoted as f$_{z}$ , f$_{x}$ and f$_{y}$ respectively. Using the example of a F$_{y}$
field the splitting of the orbital levels are given by $\pm$ \textit{d} $\sqrt{f_{x}^{2} + (f_{y} +
F_{y})^{2}}$. These quadratic variations are consistent with the observations shown in Fig 3(a). The
minimum separation of the two branches is determined by the orthogonal strain field f$_{x}$ = 1
MVm$^{-1}$ and the zero off-set is determined by the parallel strain field f$_{y}$ = 0.15 MV
m$^{-1}$.

The fine structure is changed by the electric field and in the upper branch the spin-orbit
separations are reduced as the splitting increases (Fig 4(a)). In the lower branch the state
associated with S$_{z}$ spin changes from being the highest in energy at low electric field but
crosses to be the lowest at high field. However, the non-axial spin-orbit interaction previously
neglected can cause mixing and change crossings to avoided crossings. In the case of a F$_{y}$ field
as can occur with the electric field aligned with a $<$100$>$ direction there are two avoided
crossings, (i) $\&$ (ii). Setting $\lambda$$<$E$||$L$_{xy}||$E$>$ = \textit{e} = 0.2 GHz the
calculated variation shown in Fig 4(a) is in good correspondence with the experimental situation of
Fig 3(b).

The eigenfunctions are calculated and the spin projections associated with S$_{z}$ states are shown
in Fig 4(b). The lowest trace shows the mixing of the basis states in the region of the two avoided
crossings ,(i) $\&$ (ii). The variation of mixing is consistent with the observations in the
two-laser experiment (Fig 3(b)). The situation with the upper branch is totally different. In this
case the S$_{z}$ spin level is robust to spin mixing (Fig 4(b)). For a small electric field there is
less than 1 in 10$^{3}$ component of other spins (iii) and the mixing remains small over the range of
electric fields calculated. Transitions from the S$_{z}$ ground level to this upper branch level will
give a cyclic transition and enable it to be observed in a single laser excitation spectrum. For the
other upper branch states there can be decay via a singlet so that the transitions are not cyclic and
will not be observed with one laser. Also as there is no (S$_{x}$, S$_{y}$) - S$_{z}$ mixing the
levels will not be observed in the two-laser experiments. It is concluded that model gives good
correspondence with all of the general characteristics observed experimentally.

The model allows us to comment on other issues. For example it is noted that external electric field
duplicate the effects of naturally occurring strain and the cyclic transitions reported for strained
diamond will be equivalent to those for electric field above. The cyclic transitions are optimal at
low but not zero strain (Fig 4(b)) and can occur when there is a $\Lambda$ transition associated with
the lower orbital branch. This gives a positive answers to the question of whether that cyclic and
$\Lambda$ transitions can occur simultaneously for a single NV center. By varying the field or strain
one transitions type can be optimized but the other transition is degraded. It is also possible to
anticipate the degree of spin polarization. Near an avoid crossing situation there will be almost no
spin polarization when exciting the lower branch but almost total using the upper branch, and perhaps
30$\%$ when averaging both as with vibronic excitation. When using vibronic excitation the spin
polarization can be enhance by work well away from the avoided crossing situation as can be achieved
with strain or electric field splittings greater than 15 GHz. Judging from inhomogeneous line widths
such splittings may regularly occur and with such samples high spin polarizations maybe attainable.
Other properties for ensembles can be obtained by averaging over the strain distribution of NV single
centers.

The present work has provided significant insight into the $^{3}$E excited state and when combined
with previous studies has given us an excellent understanding of the electronic structure of the NV
center. This is invaluable for the development of applications for the NV center but in particular
for its use in quantum information processing. An all optical control of single electron and probably
even single nuclear spins is within reach. This might be of particular importance for quantum memory
and repeater schemes \cite{Childress} where efficient photon to spin state conversion is of
importance. Fundamental issues and limits can be determined and it can be seen when and how the
system can be optimized for specific applications. Moreover this study demonstrates that for the NV
center the relative strength of spin flip versus spin allowed optical transitions can be tuned by an
external control parameter. This unusual property might facilitate efficient shelving and retrieval
of quantum states.

The work is support by a QAP grant from the European Commission, a SFB/TR 21 grant from DFG,
Atomoptik grant from Landesstftung BW, QuIST support from DARPA and grants from the Australian
Research Council and DSTO. Ph. T. acknowledges the Alexander von Humbolt Foundation for a research
fellowship.

\end{document}